\documentclass[a4paper,11pt]{article}
\usepackage{heppub}
\usepackage{mathtools}
\pdfoutput=1
\newcommand{\df}{\mathrm{d}}
\newcommand{\img}{\mathrm{i}}

\newcommand{\as}{\alpha_s}
\newcommand{\asL}{a_s L}
\newcommand{\nn}{\nonumber}

\newcommand{\rt}[1]{\delta_{#1}}
\newcommand{\rs}[1]{r_{#1}}
\newcommand{\rsp}[1]{r'_{#1}}
\newcommand{\barrs}[1]{\bar r_{#1}}
\newcommand{\barrsp}[1]{\bar r'_{#1}}

\newcommand{\cO}{\mathcal{O}}
\newcommand{\id}{\mathbb{I}}
\DeclareMathOperator{\Res}{Res}
\DeclareMathOperator{\Li}{Li}

\title{Analytic results for Sudakov form factors in QCD}

\author{Markus A.~Ebert}
\emailAdd{ebert@mpp.mpg.de}
\affiliation{Max Planck Institut f\"ur Physik, F\"ohringer Ring 6, 80805 Munich, Germany}

\abstract{%
Sudakov form factors appear ubiquitously in factorized cross sections
where they allow one to resum large logarithms to all orders in perturbation theory.
Their exact evaluation requires numerical integrals over anomalous dimensions,
which in practice can hamper efficiency.
Alternatively, one can use approximate analytic solutions, which provide fast evaluation
at the cost of numerical precision and loss of properties such as renormalization group invariance.
We provide an exact analytic expression of the QCD Sudakov form factor
which allows one to obtain fast and numerically exact results.
}

\date{}

\preprint{\vbox{
\hbox{MPP--2021--183}
}}

\begin{document}

\maketitle

\section{Introduction}
\label{sec:intro}

Differential observables at colliders often resolve additional collinear or soft QCD emissions
on top of the underlying Born process. Typically, their perturbative expansion then
exhibits Sudakov double logarithms of the form $\as^n \ln^m(Q/k)$, where $m \le 2n$,
$Q$ is the relevant hard scale, and $k$ is the differential observable. When the two scales
become widely separated, $k \ll Q$, these logarithms become large and can eventually spoil
the convergence of the perturbative expansion. In this case, they need to be resummed
to all orders in $\as$ to restore the reliability and stability of perturbation theory.
The resummation commonly involves Sudakov form factors of the form
\begin{align} \label{eq:def_U}
 U(\mu_0, \mu; Q) &
 = \exp\biggl\{ \int_{\mu_0}^{\mu} \frac{\df\mu'}{\mu'} \Bigl[\Gamma[\as(\mu')] \ln\frac{Q}{\mu'}
               + \gamma[\as(\mu')] \Bigr ] \biggr\}
\,,\end{align}
where $\Gamma$ is (related to) the cusp anomalous dimension and $\gamma$ is the noncusp
anomalous dimension, and their precise form depends on the observable in question.
\Eq{def_U} manifestly exponentiates and thereby resums the large logarithms
$\ln(\mu/\mu_0) \sim \ln(Q/k)$.

Sudakov form factors such as \eq{def_U} appear ubiquitously in QCD.
They were studied a long time ago in the context of
transverse momentum ($q_T$) resummation~\cite{Parisi:1979se, Curci:1979bg, Collins:1984kg},
threshold resummation~\cite{Sterman:1986aj, Catani:1989ne},
as well as event shapes~\cite{Catani:1992ua}.
In recent years, there has also been much interest in the analytic resummation of
double-differential observables~\cite{Larkoski:2014tva, Procura:2018zpn, Michel:2018hui, Lustermans:2019plv, Monni:2019yyr}
and jet-related observables such as cross sections with jet vetoes~\cite{Banfi:2012yh, Becher:2012qa, Tackmann:2012bt}.
Analytic higher-order resummation has also been utilized to achieve NNLO+PS matching~%
\cite{Alioli:2012fc, Alioli:2013hqa, Alioli:2015toa, Monni:2019whf, Lombardi:2020wju, Alioli:2021qbf}.

For a few observables, resummation has reached next-to-next-next-to-leading logarithmic (N$^3$LL) accuracy,
such as event shapes at lepton colliders~\cite{Becher:2008cf, Abbate:2010xh, Chien:2010kc, Hoang:2014wka, Ebert:2020sfi},
the $q_T$ spectrum in gluon fusion Higgs production~\cite{Chen:2018pzu, Bizon:2018foh, Billis:2021ecs, Re:2021con},
and the Drell-Yan $q_T$ spectrum~\cite{Bizon:2019zgf, Scimemi:2019cmh, Bacchetta:2019sam, Ebert:2020dfc, Becher:2020ugp, Camarda:2021ict, Re:2021con, Ju:2021lah}.
Reaching such high perturbative accuracy is particularly important for extracting the strong coupling constant
from event shapes~\cite{Becher:2008cf, Abbate:2010xh, Chien:2010kc, Hoang:2014wka}
and for determining nonperturbative transverse momentum distributions from Drell-Yan data~\cite{Scimemi:2019cmh, Bacchetta:2019sam},
as well as to match the (sub-)percent precision of $q_T$ measurements reached at the LHC
\cite{Aad:2014xaa, Aad:2015auj, Aad:2019wmn, Chatrchyan:2011wt, Khachatryan:2015oaa, Khachatryan:2016nbe, Sirunyan:2019bzr}.

When aiming for percent level accuracy, it becomes important how to evaluate \eq{def_U}.
First, there are various ways to specify the resummation accuracy,
and in general different prescriptions lead to slightly different results for the Sudakov form factor,
see e.g.~\refscite{Almeida:2014uva, Billis:2019evv}.
However, such differences formally constitute a subleading effect and should be well covered
by any reliable estimate of perturbative uncertainties, and hence are not the focus of this paper.
Instead, we focus on numerically precise evaluations of the integral in \eq{def_U}.
A priori, this seems to be a straightforward task, as one can numerically evaluate the integral
at the desired precision. However, such numerical integrals may have significant impact on
the runtime of computer codes, as they typically appear in the hottest loop.
In particular, this can become a concern when the running coupling $\as(\mu')$
is to be evaluated numerically exact, which beyond LL is not possible in closed form.

In practice, most phenomenological studies employ an analytic approximation
of the integral in \eq{def_U} that neglects higher-order terms
beyond the desired formal resummation accuracy. Despite retaining formal accuracy,
such analytic approximations can break
important properties of the Sudakov form factor. For example, \refcite{Bell:2018gce}
explicitly discussed the breaking of the relation
\begin{align} \label{eq:intro_RG_invariance}
 U(\mu_0, \bar\mu; Q) U(\bar\mu, \mu; Q) = U(\mu_0, \mu; Q)
\,,\end{align}
which holds for any $\bar\mu$ as a consequence of \eq{def_U}
and encodes RG invariance of the cross section.
\Refcite{Bell:2018gce} traced back the violation
of \eq{intro_RG_invariance} to the relation
\begin{align} \label{eq:beta}
 \frac{\df\as(\mu)}{\df\ln\mu} = \beta[\as(\mu)]
\qquad\Rightarrow\qquad
 \ln\frac{\mu}{\mu_0} = \int_{\as(\mu_0)}^{\as(\mu)} \frac{\df \as}{\beta(\as)}
\,,\end{align}
where $\beta(\as)$ is the QCD $\beta$ function.
\Eq{beta} is broken by terms beyond the formal resummation accuracy
when an approximate solution for the running coupling constant $\as(\mu)$ is used.
This effect leads to an undesired dependency on the (formally arbitrary) scale $\bar\mu$
when combining multiple Sudakov form factors.

A direct consequence of \eq{intro_RG_invariance} is the
\emph{closure condition},
\begin{align} \label{eq:intro_closure}
 U(\mu_0, \bar\mu; Q) U(\bar\mu, \mu_0; Q) = 1
\,,\end{align}
which was studied numerically in \refcite{Billis:2019evv}
in the context of coupled gauge theories.
There, the authors found that commonly employed analytic approximations of $U$
violate \eq{intro_closure} at the level of up to a few percent,
which can be a prohibitively large effect at N$^3$LL accuracy.
In \refcite{Bozzi:2020}, it was even argued that it may become necessary
to consider the non-closure under RG invariance, there dubbed ``perturbative hysteresis'',
as an additional theory uncertainty.
As a direct solution, \refcite{Billis:2019evv} proposed a seminumerical method,
where the integral in \eq{def_U} is evaluated numerically
in conjunction with an analytic approximation of the running coupling $\as(\mu')$,
which by construction obeys \eqs{intro_RG_invariance}{intro_closure}.

\Refcite{Billis:2019evv} also provided a detailed numerical comparison
of several approximate solutions to the numerically exact evaluation of \eq{def_U}.
While their seminumerical approach yielded errors of only $\cO(0.1\%)$
and thus provides a very good approximation to the exact form factor,
commonly employed approximations led to unacceptable discrepancies of up to $\cO(10\%)$.

In this paper, we overcome the above difficulties by deriving an \emph{exact}
analytic solution of \eq{def_U}. This allows one to obtain exact results
at much better numerical efficiency than numeric integration methods.
In particular, our result trivially obeys RG invariance
and can, in principle, be evaluated at arbitrary precision.

This paper is structured as follows. In \sec{Sudakov},
we discuss the general structure of the form factor
and define the logarithmic order counting used to the specify
the perturbative accuracy of our results. In \sec{anal_solutions},
we derive the analytic solution for the Sudakov form factor, before concluding in
\sec{conclusions}. Useful reference formulas are provided in \app{results}.

\section{The Sudakov form factor}
\label{sec:Sudakov}

We first rewrite the Sudakov form factor in \eq{def_U} as
\begin{align} \label{eq:def_U_2}
 U(\mu_0, \mu; Q) &
 = \exp\biggl\{ \int_{\mu_0}^{\mu} \frac{\df\mu'}{\mu'} \Bigl[\Gamma[\as(\mu')] \ln\frac{Q}{\mu'}
               + \gamma[\as(\mu')] \Bigr ] \biggr\}
\\\nn&
 = \exp\biggl\{ - K_\Gamma[\as(\mu_0), \as(\mu)]
                + \ln\frac{Q}{\mu_0} \eta_\Gamma[\as(\mu_0), \as(\mu)]
                + K_\gamma[\as(\mu_0), \as(\mu)] \biggr\}
\,,\end{align}
where the following basic kernels will be the focus of our study:
\begin{alignat}{2} \label{eq:kernel_defs}
 K_\Gamma[\as(\mu_0), \as(\mu)] &
 = \int_{\mu_0}^\mu \frac{\df\mu'}{\mu'} \Gamma[\as(\mu')] \ln\frac{\mu'}{\mu_0}
 &&= \int_{\as(\mu_0)}^{\as(\mu)} \frac{\df \as}{\beta(\as)} \Gamma(\as)
   \int_{\as(\mu_0)}^{\as} \frac{\df\as'}{\beta(\as')}
\,,\nn\\
 K_\gamma[\as(\mu_0), \as(\mu)] &
 = \int_{\mu_0}^\mu \frac{\df\mu'}{\mu'} \gamma[\as(\mu')]
 &&= \int_{\as(\mu_0)}^{\as(\mu)} \frac{\df \as}{\beta(\as)} \gamma(\as)
\,,\nn\\
 \eta_\Gamma[\as(\mu_0), \as(\mu)] &
 = \int_{\mu_0}^\mu \frac{\df\mu'}{\mu'} \Gamma[\as(\mu')]
 &&= \int_{\as(\mu_0)}^{\as(\mu)} \frac{\df \as}{\beta(\as)} \Gamma(\as)
\,.\end{alignat}
Here, the second form is obtained by employing \eq{beta}.
It makes clear that $K_\Gamma, K_\gamma$ and $\eta_\Gamma$ depend on the scales $\mu_0$
and $\mu$ only through $\as(\mu_0)$ and $\as(\mu)$, respectively.
It also has the practical advantage that the running coupling
has to be evaluated only at the external scales rather than continuously inside the integral,
which is particularly useful when it is obtained as external input, e.g.~from
the parton distribution functions, or by an exact numeric solution of \eq{beta}.

The second form in \eq{kernel_defs} also opens a straightforward way to obtain
an analytic approximation by expanding the integrands in $\as$ to the desired order,
after which one can straightforwardly evaluate the integrals over $\as$.
For more details and explicit results up to N$^3$LL, see e.g.~\refcite{Billis:2019evv}.

\subsection{Resummation accuracy}
\label{sec:accuracy}

The goal of this paper is to obtain exact analytic formulas for the integrals
in \eq{kernel_defs}. For this purpose, it is important to \emph{precisely} define
the perturbative accuracy with which to evaluate the integrals, i.e.~their formal resummation accuracy.

We follow the common definition of classifying the perturbative accuracy of \eq{kernel_defs}
through the perturbative order to which the cusp anomalous dimension $\Gamma(\as)$,
the noncusp anomalous dimension $\gamma(\as)$, and the $\beta$ function are kept.
Their fixed-order expansions read
\begin{align}
 \Gamma(\as) = \sum_{k=0}^\infty \Gamma_k \Bigl(\frac{\as}{4\pi}\Bigr)^{k+1}
\,,\quad
 \gamma(\as) = \sum_{k=0}^\infty \gamma_k \Bigl(\frac{\as}{4\pi}\Bigr)^{k+1}
\,,\quad
 \beta(\as) = -2 \as \sum_{k=0}^\infty \beta_k \Bigl(\frac{\as}{4\pi}\Bigr)^{k+1}
\,.\end{align}
N$^n$LL accuracy is then specified by truncating $\Gamma(\as)$ at $\cO(\as^{n+1})$,
i.e.~at $n{+}1$ loops, while $\gamma(\as)$ enters at one lower order, i.e.~at $\cO(\as^{n})$.
The running coupling is also evaluated at exactly $n{+}1$-loops,
i.e.~the beta function is calculated at $\cO(\as^{n+2})$.
This convention is summarized in table~\ref{tbl:NnLL_counting}.
Note that the different orders at which $\Gamma$ and $\gamma$ enter the
resummed form factor are the reason to treat $K_\gamma$ and $\eta_\Gamma$
separately, even though their formal (all-order) definitions
in \eq{kernel_defs} are identical.

In order to illustrate the importance of a precise definition of the resummation accuracy,
we note that one can expand the Sudakov form factor as
\begin{align} \label{eq:Sud_schematic}
 U(\mu_0 = k, \mu = Q; Q)
 = \exp\bigl[ L \, g_{\rm LL}(\asL) + g_{\rm NLL}(\asL) + a_s \, g_{\rm NNLL}(\asL) + \cdots \bigr]
\,,\end{align}
where for simplicity we chose $\mu_0 = k$  and $\mu=Q$,
and defined the abbreviations $a_s = \as(Q) / (4\pi)$ and $L = \ln(Q/k)$.
The first few terms of this expansion read
\begin{alignat}{3} \label{eq:g_res}
 g_{\rm LL}(\asL) &
 = \frac{\Gamma_0}{2} \asL &&~+~ (\asL)^2 \frac23 \Gamma_0 \beta_0  &&~+~ \cO[(\asL)^3]
\,,\nn\\
 g_{\rm NLL}(\asL) &
 = \gamma_0 \asL &&~+~ (\asL)^2 \Bigl(\frac{\Gamma_1}{2} + \gamma_0 \beta_0 \Bigr) &&~+~ \cO[(\asL)^3]
\,,\nn\\
 g_{\rm NNLL}(\asL) &
 = \gamma_1 \asL &&~+~ (\asL)^2 \Bigl(\frac{\Gamma_2}{2} + 2 \gamma_1 \beta_0 + \gamma_0 \beta_1 \Bigr) &&~+~ \cO[(\asL)^3]
\,,\end{alignat}
where we have expanded \eq{g_res} in $\as L \ll 1$ to obtain more compact expressions.
In practice, one counts $\as L \sim 1$ and such an expansion is not valid.
The corresponding unexpanded expressions can be found e.g.~in \refcite{deFlorian:2004mp},
up to different overall conventions.

\Eqs{Sud_schematic}{g_res} show that in order to exponentiate
all leading logarithms, i.e.~retaining all terms of $\cO(\as^n L^{n+1})$ \emph{in the exponent},
one must at least keep $\Gamma_0$ and $\beta_0$ exact. At NLL, one also needs to keep at least
$\Gamma_{1}$, $\gamma_0$ and (not shown in \eq{g_res}) $\beta_{1}$.
Extending this to higher orders results in the classification in table~\ref{tbl:NnLL_counting}.
From \eq{g_res}, one can also see that formally lower-order terms also appear in the
higher-order functions. For example, the term proportional to $\gamma_0 \beta_1$ arising
in $g_{\rm NNLL}$ is classified as a NNLL term, but is already induced when keeping
the dependence on $\gamma_0$ and $\beta_1$ exact as specified by our NLL counting.

In principle, one may define N$^n$LL by keeping only the corresponding terms $g_{\text{N}^n\text{LL}}$
in \eq{Sud_schematic}. As illustrated, this differs from our classification of
evaluating each anomalous dimension in \eq{def_U} according to table~\ref{tbl:NnLL_counting}.
Both definitions are well defined, and they only differ by higher-order terms in either counting.
In practice, as long as the difference between such prescriptions
is well covered by an estimate of theory uncertainties, either approach is justified.
However, prescriptions truncating the exponent as in \eq{Sud_schematic}
naturally break the RG invariance (except at LL and N$^\infty$LL),
i.e.~\eq{intro_RG_invariance}, raising the question whether
such a shortcoming should be assigned its own uncertainty~\cite{Bozzi:2020}.
Thus, from a purely theoretical point of view it is advisable
to classify the resummation accuracy solely via the integrand in \eq{def_U}.
The importance of formally defining the logarithmic order-counting
via the fixed-order perturbative expansion of the anomalous dimensions,
as opposed to retaining towers of logarithms in the solution of the Sudakov,
was also highlighted in \refscite{Almeida:2014uva, Ebert:2016gcn}.

{
\renewcommand{\arraystretch}{1.2}
\begin{table*}[pt]
 \centering
 \begin{tabular}{l|c|c|c} \hline\hline
  Accuracy     & $\Gamma(\as)$ & $\gamma(\as)$ & $\beta(\as)/\as$ \\\hline
  LL           & $\cO(\as^1)$   & --            & $\cO(\as^1)$ \\\hline
  NLL          & $\cO(\as^2)$   & $\cO(\as^1)$  & $\cO(\as^2)$ \\\hline
  NNLL         & $\cO(\as^3)$   & $\cO(\as^2)$  & $\cO(\as^3)$ \\\hline
  N$^3$LL      & $\cO(\as^4)$   & $\cO(\as^3)$  & $\cO(\as^4)$ \\\hline
 \hline
 \end{tabular}
 \caption{Classification of the resummation accuracy in terms of the
          fixed-order expansions of the anomalous dimensions and beta function.}
 \label{tbl:NnLL_counting}
\end{table*}
\renewcommand{\arraystretch}{1.0}
}

\subsection{The role of the running coupling in retaining RG invariance}
\label{sec:rg_invariance}

In \sec{accuracy}, we specified that for exact N$^n$LL accuracy the QCD $\beta$
function is solved at exactly $n{+1}$ loop order.
In practice, this may not always be feasible. For example, for consistency
with external input such as parton distribution functions, one may wish
to take the running coupling as external input itself, in which case
it may not obey the exact $\beta$ function due to formally higher-order terms.
The same happens if one evaluates the running coupling at higher logarithmic
accuracy than in the resummation itself. Both cases are formally well defined
as long as they differ from strict N$^n$LL accuracy only by higher-order terms
in the $\beta$ function.

Nevertheless, in such a case the two forms in \eq{kernel_defs} are not equivalent any more,
and one must choose which one to define as the fundamental quantity.
For the practical reasons mentioned before, we define the kernels through
the second form in \eq{kernel_defs}, i.e.~as integrals over $\as$ rather than $\mu'$.
However, in this case one breaks the RG invariance, i.e.~\eq{intro_RG_invariance},
as already studied in detail in \refcite{Bell:2018gce}.
Since the RG invariance is only broken beyond the formal logarithmic accuracy,
this procedure still defines a valid scheme.

Of course, it is still desirable to retain RG invariance as specified by \eq{intro_RG_invariance}.
Using \eq{def_U_2}, we can rewrite this condition as
\begin{align} \label{eq:consistency_Gamma}
 & K_\Gamma[\as(\mu_0), \as(\mu)] - K_\Gamma[\as(\mu_0), \as(\bar\mu)] - K_\Gamma[\as(\bar\mu), \as(\mu)]
 \nn\\
 =&~ \ln\frac{Q}{\mu_0} \eta_\Gamma[\as(\mu_0), \as(\mu)]
 - \ln\frac{Q}{\mu_0} \eta_\Gamma[\as(\mu_0), \as(\bar\mu)] - \ln\frac{Q}{\bar\mu} \eta_\Gamma[\as(\bar\mu), \as(\mu)]
\end{align}
for the cusp piece, and
\begin{align} \label{eq:consistency_gamma}
 K_\gamma[\as(\mu_0), \as(\mu)] = K_\gamma[\as(\mu_0), \as(\bar\mu)] + K_\gamma[\as(\bar\mu), \as(\mu)]
\end{align}
for the noncusp piece. The latter is trivially true for either form in \eq{kernel_defs}.
To validate \eq{consistency_Gamma}, we take the derivative with respect to
${\bar\alpha}_s \equiv \as(\bar\mu)$, which yields
\begin{align} \label{eq:consistency_Gamma_2}
 & - \frac{\Gamma({\bar\alpha}_s)}{\beta({\bar\alpha}_s)} \int_{\as(\mu_0)}^{{\bar\alpha}_s} \frac{\df\as'}{\beta(\as')}
   + \int_{{\bar\alpha}_s}^{\as(\mu)} \frac{\df \as}{\beta(\as)} \frac{\Gamma(\as)}{\beta({\bar\alpha}_s)}
 \nn\\
 =&~ - \ln\frac{Q}{\mu_0} \frac{\Gamma({\bar\alpha}_s)}{\beta({\bar\alpha}_s)}
      + \frac{1}{\beta({\bar\alpha}_s)}\int_{{\bar\alpha}_s}^{\as(\mu)} \frac{\df \as}{\beta(\as)} \Gamma(\as)
      + \ln\frac{Q}{\bar\mu} \frac{\Gamma({\bar\alpha}_s)}{\beta({\bar\alpha}_s)}
\,,\end{align}
which can be simplified to
\begin{align} \label{eq:consistency_Gamma_3}
 \int_{\as(\mu_0)}^{{\bar\alpha}_s} \frac{\df\as'}{\beta(\as')}
 = \ln\frac{\bar\mu}{\mu_0}
\,.\end{align}
As expected, \eq{consistency_Gamma} is true if and only if the QCD $\beta$ function is obeyed exactly.
Conversely, using a running coupling $\as(\mu)$ which does not exactly obey the $\beta$ function
at the specified order will violate RG invariance~\cite{Bell:2018gce}.

A possible solution to this problem was already found in \refcite{Bell:2018gce}
by imposing \eq{consistency_Gamma_3} at the level of the Sudakov form factor.
Specifically, we rewrite \eq{def_U_2} as
\begin{align} \label{eq:def_U_3}
 U(\mu_0, \mu; Q)
 = \exp\Bigl\{&\!- K_\Gamma[\as(\mu_0), \as(\mu)]
                + L[\as(\mu_0), \as(Q)] \, \eta_\Gamma[\as(\mu_0), \as(\mu)]
                \nn\\&
                + K_\gamma[\as(\mu_0), \as(\mu)] \Biggr\}
\,,\end{align}
where we replaced $\ln(Q/\mu_0)$ by%
\footnote{One may be tempted to write this integral as $\eta_\id(\mu_0,Q)$ or $K_\id(\mu_0,Q)$ with
$\id(\as) = 1$, but the corresponding analytic expressions are not applicable
as they are derived assuming $\id(\as) = \cO(\as)$.}
\begin{align} \label{eq:def_L}
 L[\as(\mu_0), \as(Q)] = \int_{\as(\mu_0)}^{\as(Q)} \frac{\df \as}{\beta(\as)}
 \qquad\xrightarrow[~\text{running}]{\text{exact}~\as}\qquad
 \ln\frac{Q}{\mu_0}
\,.\end{align}
For exact $\as$ running, this is by definition identical to $\ln(Q/\mu_0)$,
while otherwise it differs from it by formally subleading terms.
Thus, \eq{def_U_3} constitutes a valid definition of resummation accuracy on its own.

\section{Analytic evaluation of the Sudakov form factor}
\label{sec:anal_solutions}

In this section, we obtain analytic results for the basic building blocks
$K_\Gamma, K_\gamma$ and $\eta_\Gamma$ defined in \eq{kernel_defs},
which in turn yield an exact solution of the Sudakov form factor in \eq{def_U}.
The key ingredient is the generic indefinite integral over a rational function
of the form $f(\as) / \beta(\as)$, which is discussed in detail in \sec{indef_integral}.
The resulting solutions for $\eta_\Gamma$ and $K_\gamma$ are presented in \sec{sol_etaGamma_Kgamma},
while the result for $K_\Gamma$ is more involved and presented in \sec{sol_KGamma}.
We also briefly comment on the option to numerically solve the QCD $\beta$ function
with our method in \sec{sol_beta}.

The results presented in this section have been validated
against numerical implementations of the RG kernels,
and were already used for the N$^3$LL$^\prime$ resummation of the
Energy-Energy Correlation in \refcite{Ebert:2020sfi} based on an
implementation in \texttt{SCETlib}~\cite{scetlib}.

\subsection{\texorpdfstring{Indefinite integral over $f(\as)/\beta(\as)$}
                           {Indefinite integral over f(as)/beta(as)}}
\label{sec:indef_integral}

We start by considering the generic indefinite integral
\begin{align} \label{eq:beta_int_0}
 I_f(\as) \equiv \int\df \as \frac{f(\as)}{\beta(\as)}
\,,\end{align}
where $f(\as)$ and $\beta(\as)$ are polynomials in $\as$,
expanded in $a_s \equiv \as / (4\pi)$ as
\begin{align} \label{eq:expansion}
 f(\as) = \sum_{k=-1}^m f_k a_s^{k+1}
\,,\qquad
 \beta(\as) = -8 \pi a_s \sum_{k=0}^n \beta_k a_s^{k+1}
\,.\end{align}
While anomalous dimensions obey  $f(\as) = \cO(\as)$,
it will be useful to allow for a nonvanishing constant $f_{-1} \ne 0$
to also cover the case of $L(\mu_0,Q)$ defined in \eq{def_L}.
The expansions in \eq{expansion} are truncated at $\cO(\as^m)$ and $\cO(\as^n)$,
respectively, with $m$ and $n$ given according to table~\ref{tbl:NnLL_counting}.

Our definition of N$^n$LL accuracy implies that
\begin{align} \label{eq:beta_int_1}
 m \le n
 \qquad\Leftrightarrow\qquad
 {\rm deg}(f) < {\rm deg}(\beta)
\,,\end{align}
such that $\beta(\as)$ is always a polynomial of higher degree than $f(\as)$.
It follows that $f(\as)/\beta(\as)$ is a proper rational function,
and as such its partial fraction decomposition has no finite remainder.
To obtain it, we first factorize the $\beta$ function as
\begin{align} \label{eq:beta_int_2}
 \beta(\as)
 = -8\pi a_s^2 (\beta_0 + \beta_1 a_s + \cdots + \beta_n a_s^n)
 = -8\pi a_s^2 \beta_n \prod_{i=1}^n \bigl(a_s - \rt{i} \bigr)
\,,\end{align}
where $n$ is the truncation order specified by \eq{expansion},
and we factored out the highest coefficient $\beta_n$.
The $\rt{i}$ are the corresponding nonvanishing roots of $\beta(\as)$,
which implicitly depend on the order $n$ of the $\beta$ function expansion.
For later convenience, we also define $\rt{0} \equiv 0$.
Recall that the roots of a polynomial are in general complex, even though $\beta(\as)$ itself is a real function.
For $\beta(\as)$, starting at NNLL ($n=2$) one indeed encounters complex roots,
such that all following expressions are to be understood within complex analysis.
Explicit expressions for the $\rt{i}$ up to N$^3$LL are provided in \app{beta_roots}.

We now write the partial fraction decomposition of $f(\as)/\beta(\as)$ as
\begin{align} \label{eq:beta_int_4}
 \frac{f(\as)}{\beta(\as)}
 = -\frac{1}{8\pi} \biggl[ \frac{\rsp{f,0}}{a_s^2} + \frac{\rs{f,0}}{a_s} + \sum_{i=1}^n \frac{\rs{f,i}}{a_s - \rt{i}} \biggr]
\,,\end{align}
which has no finite or purely polynomial contributions thanks to \eq{beta_int_1}.
We also assume that all $\rt{i}$ are distinct, i.e.~$a_s = 0$ is the only root
of higher multiplicity. Using the five-loop result in \refscite{Baikov:2016tgj, Herzog:2017ohr},
one can explicitly check that this is fulfilled for the QCD $\beta$ function at least up to N$^4$LL.
In \eq{beta_int_4}, we have extracted an overall factor $1/(8\pi)$ to obtain more
compact expressions for the coefficients $\rs{f,i}$.
Note that the double pole $\rsp{f,0} / a_s^2$ vanishes if $f(\as) = \cO(\as)$,
as is the case for anomalous dimensions.

The coefficients $\rsp{f,0}$ and $\rs{f,i}$ can be regarded as the coefficient
of the corresponding pole of the Laurent series around $\rt{i}$,
and thus are given in terms of the corresponding residues,
\begin{align}
 \rsp{f,0} &= -8\pi \Res_{a_s = 0}\biggl[a_s \frac{f(4\pi a_s)}{\beta(4\pi a_s)}\biggr]
\,,\quad
 \rs{f,i} = -8\pi \Res_{a_s = \rt{i}} \biggl[\frac{f(4\pi a_s)}{\beta(4\pi a_s)}\biggr]
\,.\end{align}
Note that these coefficients depend on the order $n$ of the $\beta$ function
and its roots, and thus differ at different resummation orders.
For simplicity, we keep this dependence implicit.
The residues can be explicitly evaluated in terms of the coefficients in \eqs{expansion}{beta_int_2},
\begin{align} \label{eq:kappa}
 \rsp{f,0} = \frac{f_{-1}}{\beta_0}
\,,\qquad
 \rs{f,0} = \frac{f_0}{\beta_0} - \theta_{n>0} \frac{f_{-1} \beta_1}{\beta_0^2}
\,,\qquad
 \rs{f,i>0}= -2 \frac{f(4\pi\rt{i})}{\beta'(4\pi\rt{i})}
\,.\end{align}
Here, the $\theta_{n>0}$ is necessary, as for $n=0$ one has $\beta(\as) = -8 \pi a_s^2$,
and thus a single pole from $\beta_1$ is not yet allowed.

Using \eq{beta_int_4}, it is trivial to evaluate the indefinite integral in \eq{beta_int_0},
\begin{align} \label{eq:beta_int_5}
 I_f(\as)
 = \int\df \as \frac{f(\as)}{\beta(\as)}
 = \frac{\rsp{f,0}}{2 a_s} - \frac{\rs{f,0} }{2} \ln a_s
    - \frac{1}{2} \sum_{i=1}^n \rs{f,i} \ln\bigl(a_s - \rt{i}\bigr)
\,.\end{align}
The simple expressions in \eqs{kappa}{beta_int_5} are key ingredients
to analytically evaluate the integrals in \eq{kernel_defs}.

Before proceeding, we briefly discuss the case of complex roots $\rt{i}$.
While from \eq{kappa} it is clear that $\rsp{f,0}$ and $\rs{f,0}$ are manifestly real,
the $\rs{f,i}$ and $\rt{i}$ are in general complex.
However, by the complex conjugate root theorem,
if $\rt{i}$ is a complex root with coefficient $\rs{f,i}$,
then its conjugate $\rt{i}^*$ is also a root with the conjugate coefficient $\rs{f,i}^*$.
The contribution of such a pair of complex roots to \eq{beta_int_4} is given by
\begin{align}
 \frac{\rs{f,i}}{a_s - \rt{i}} + \frac{\rs{f,i}^*}{a_s - \rt{i}^*} &
 = 2 \frac{a_s \Re(\rs{f,i}) - \Re(\rs{f,i} \rt{i}^*)}{[a_s - \Re(\rt{i})]^2 + \Im(\rt{i})^2}
\,,\end{align}
where $\Re(x)$ and $\Im(x)$ denote the real and imaginary part of $x$, respectively.
The indefinite integral then becomes
\begin{align} \label{eq:beta_int_7}
 I_f(\as) &
 = \frac{\rsp{f,0}}{2a_s}  -  \sum_{i=0}^n \biggl\{
     \frac{\Re(\rs{f,i})}{4} \ln\bigl[(a_s - \Re(\rt{i}))^2 + \Im(\rt{i})^2\bigr]
     - \frac{\Im(\rs{f,i})}{2} {\rm atan}\biggl[\frac{a_s - \Re(\rt{i})}{\Im(\rt{i})}\biggr] \biggr\}
\,,\end{align}
where we treat real and complex roots on equal footing.
If desired, one can thus write \eq{beta_int_5} in a manifestly real fashion
to avoid complex-valued logarithms.
\Eq{beta_int_7} trivially reproduces \eq{beta_int_5}
in the special case that all imaginary parts vanish.
In the following, we will always employ \eq{beta_int_5},
as it leads to more compact expressions than \eq{beta_int_7}.

\subsection{Analytic solutions for \texorpdfstring{$\eta_\Gamma$ and $K_\gamma$}{etaGamma and Kgamma}}
\label{sec:sol_etaGamma_Kgamma}

From \eq{kernel_defs}, it is obvious that $\eta_\Gamma$ and $K_\gamma$
can be trivially obtained from \eq{beta_int_5}. Since they involve
anomalous dimensions starting at $\cO(\as)$, we have $\rsp{f,0} = 0$,
such that we obtain
\begin{align} \label{eq:etaGamma_Kgamma}
 \eta_\Gamma(\alpha_0, \as) &
 = - \frac{1}{2} \sum_{i=0}^n \rs{\Gamma,i} \ln\frac{\as- 4\pi\rt{i}}{\alpha_0  - 4\pi \rt{i}}
\,,\nn\\
 K_\gamma(\alpha_0, \as) &
 = - \frac{1}{2} \sum_{i=0}^n \rs{\gamma,i} \ln\frac{\as - 4\pi\rt{i}}{\alpha_0 - 4\pi \rt{i}}
\,.\end{align}
We stress again that the residues $\rs{f,i}$ and the roots $\rt{i}$
depend on the truncation order $n$. From \eq{kappa}, we have
\begin{alignat}{2} \label{eq:kappa_gamma}
 \rs{\Gamma,0}  &= \frac{\Gamma_0}{\beta_0}
\,,\qquad
 &&\rs{\Gamma,i}  = -2 \frac{\Gamma(4\pi\rt{i})}{\beta'(4\pi\rt{i})}
   \,,\qquad
   i = 1, \dots, n
\,,\nn\\
 \rs{\gamma,0}  &= \frac{\gamma_0}{\beta_0} \theta_{n>0}
\,,\qquad
 &&\rs{\gamma,i}  = -2 \frac{\gamma(4\pi\rt{i})}{\beta'(4\pi\rt{i})}
   \,,\qquad
   i = 1, \dots, n
\,,\end{alignat}
where it is important to recall that according to the counting in
table~\ref{tbl:NnLL_counting}, $\gamma$ enters at one lower order
than $\Gamma$.

\subsection{Analytic solution for \texorpdfstring{$K_\Gamma$}{KGamma}}
\label{sec:sol_KGamma}

Recall the definition of $K_\Gamma$ in \eq{kernel_defs},
\begin{align} \label{eq:beta_int_20}
 K_\Gamma(\alpha_0, \as) &
 = \int_{\alpha_0}^{\as} \df \as' \frac{\Gamma(\as')}{\beta(\as')}
   \int_{\alpha_0}^{\as'} \frac{\df\as''}{\beta(\as'')}
\,.\end{align}
The inner integral follows from \eq{beta_int_5} by setting $f(\as) \equiv 1$,
\begin{align} \label{eq:sol_L}
 L(\alpha_0, \as') &
 = \int_{\alpha_0}^{\as'} \frac{\df \as''}{\beta(\as'')}
 = \frac{\barrsp{0}}{2} \left(\frac{1}{a_s'} - \frac{1}{a_0}\right)
    - \frac{1}{2} \barrs{0} \ln\frac{a_s'}{a_0}
    - \frac{1}{2} \sum_{i=1}^n \barrs{i} \ln\frac{a_s' - \rt{i}}{a_0 - \rt{i}}
\,,\end{align}
where as before $a'_s = \as' / (4\pi)$ and $a_0 = \alpha_0 / (4\pi)$,
and the $\barrs{i}$ follow from \eq{kappa} as
\begin{align} \label{eq:bar_kappa}
 \barrsp{0} = \frac{1}{\beta_0}
\,,\qquad
 \barrs{0} = - \frac{\beta_1}{\beta_0^2} \theta_{n>0}
\,,\qquad
 \barrs{i} = \frac{-2}{\beta'(4\pi\rt{i})}
\,.\end{align}
Inserting \eq{sol_L} into \eq{beta_int_20} and using the partial
fractioning in \eq{beta_int_4}, one obtains
\begin{align} \label{eq:beta_int_22}
 K_\Gamma(\alpha_0, \as)
 &= \int_{\alpha_0}^{\as} \df \as' \frac{\Gamma(\as')}{\beta(\as')} L(\alpha_0, \as')
\\\nn&
 = -\frac{1}{4} \int_{a_0}^{a_s} \df a_s' \,
   \biggl[ \sum_{i=0}^n \frac{\rs{\Gamma,i}}{a'_s - \rt{i}} \biggr]
   \biggl[ \barrsp{0} \Bigl(\frac{1}{a'_s}-\frac{1}{a_0}\Bigr)
          - \sum_{j=0}^n \barrs{j} \ln\frac{a'_s - \rt{j}}{a_0 - \rt{j}} \biggr]
\,,\end{align}
where the $\rs{\Gamma,i}$ are the same as in \eq{kappa_gamma}.
Evaluating the remaining integral, we arrive at the final result
\begin{align} \label{eq:beta_int_23}
 K_\Gamma(\alpha_0, \as) &
 = \frac14 \rs{\Gamma,0} \barrsp{0} \Bigl( \frac{1}{a_s} + \frac{L_0-1}{a_0} \Bigr)
   + \frac18 \rs{\Gamma,0} \barrs{0} L_0^2
 \nn\\&\quad
  + \frac{1}{4} \sum_{i=1}^n \bigl(\rs{\Gamma,i} \barrs{0}  - \rs{\Gamma,0} \barrs{i}\bigr)
    \biggl[ L_0\ln\Bigl(1 - \frac{a_0}{\rt{i}}\Bigr)
           + \Li_2\Bigl(\frac{a_s}{\rt{i}}\Bigr) - \Li_2\Bigl(\frac{a_0}{\rt{i}}\Bigr) \biggr]
\nn\\&\quad
  + \frac{1}{4} \sum_{i=1}^n \rs{\Gamma,i} \biggl[
    \frac{1}{2} \barrs{i} L_i^2 + \barrs{0} L_0 L_i + \barrsp{0} \Bigl( \frac{L_0 - L_i}{\rt{i}} + \frac{L_i}{a_0} \Bigr) \biggr]
\nn\\&\quad
  + \frac{1}{4} \sum_{\substack{i,j=1\\i \ne j}}^n \rs{\Gamma,i} \barrs{j}
    \biggl[ L_j \ln\frac{\rt{i} - a_s}{\rt{i} - \rt{j}}
            + \Li_2\Bigl(\frac{a_s - \rt{j}}{\rt{i} - \rt{j}}\Bigr)
            - \Li_2\Bigl(\frac{a_0 - \rt{j}}{\rt{i} - \rt{j}}\Bigr) \biggr]
\,,\end{align}
where we defined the logarithms (again suppressing the dependence on $n$)
\begin{align}
 L_i = \ln\frac{a_s - \rt{i}}{a_0 - \rt{i}}
\,.\end{align}
Compared to \eq{etaGamma_Kgamma}, \eq{beta_int_23} is significantly more involved.
In particular, starting at NNLL one encounters complex roots and residues,
and hence the evaluation of \eq{beta_int_23} involves complex dilogarithms.
However, these are readily available in most programming languages,
and thus do not pose a problem in practice.

\subsection{Numerical solution of \texorpdfstring{$\as$}{as} running}
\label{sec:sol_beta}

As a side remark, we note that the above results can also be used for a
numerical solution of the QCD $\beta$ function.
From \eq{sol_L}, we have
\begin{align} \label{eq:beta_solver}
 \ln\frac{\mu}{\mu_0} &
 = L[\as(\mu_0), \as(\mu)]
 =  \frac{2 \pi}{\beta_0} \biggl[ \frac{1}{\as(\mu)} - \frac{1}{\as(\mu_0)} \biggr]
   - \frac{1}{2} \sum_{i=0}^n \barrs{i} \ln\frac{\as(\mu) - 4\pi \rt{i}}{\as(\mu_0) - 4\pi \rt{i}}
\,,\end{align}
where we already used that $\barrsp{0} = 1 / \beta_0$,
with the remaining $\barrs{i}$ given in \eq{bar_kappa}.
The running coupling constant $\as(\mu)$ follows by numerically
inverting \eq{beta_solver}, for example using a root finding algorithm
such as Newton's method. In particular, using an analytic approximation
of the exact running coupling as the starting point of the root finder
greatly improves its convergence. This provides an alternative approach
to established methods such as numerically solving the differential
equation, for example using the Runge-Kutta method.

\section{Conclusions}
\label{sec:conclusions}

We have presented an analytic and exact evaluation of the
Sudakov form factor that appears ubiquitously in resummed predictions.
Our results are based on a partial fraction decomposition of rational functions
of the form $f(\as) / \beta(\as)$, where $f(\as)$ is any anomalous dimension
and $\beta(\as)$ is the QCD $\beta$ function. This allows one to analytically
compute the integral with respect to $\as$, and explicit expressions for the basic
building block of the Sudakov form factor are provided.
As a byproduct, we also obtained a method to compute the running coupling
constant by means of a numeric root solver.

A key ingredient for our solution is the precise definition of logarithmic accuracy,
which we define by the perturbative order of the anomalous dimensions and $\beta$
function. This definition can not be fulfilled exactly when using a running coupling
constant $\as(\mu)$ that violates the $\beta$ function by terms beyond formal accuracy.
In this case, it is well known that renormalization group invariance can be violated,
but it can be restored using a modified definition of the Sudakov form factor
which reproduces the standard form in the case of exact running coupling.

Our results are particularly useful for resummed calculations at high accuracy.
An important example is $q_T$ resummation, where $\cO(1\%)$ accuracy is desired
and where the resummation is commonly performed prior to a Fourier and phase-space integral,
such that a fast and precise evaluation of the Sudakov form factor is indispensable.
Thanks to their analytic nature, our results can be evaluated at arbitrary precision,
and thus can also serve as a well-defined reference result to numerically validate
the accuracy of commonly used approximations. It may also prove useful
when comparing different resummation codes against each other,
as our results can remove one possible source of numerical differences.

\acknowledgments
We thank Johannes Michel and Frank Tackmann for useful comments on this manuscript,
and Chris Lee for interesting discussions concerning \refcite{Bell:2018gce}.

\appendix

\section{Coefficients of the partial fraction decomposition}
\label{app:results}

In this appendix, we summarize and provide explicit expressions for the coefficients
of the partial fraction decompositions. Following table~\ref{tbl:NnLL_counting},
at N$^n$LL accuracy, we expand the cusp and noncusp anomalous dimensions as
\begin{align} \label{eq:anom_dims}
 \Gamma(\as) = \sum_{k=0}^n \Gamma_k \Bigl(\frac{\as}{4\pi}\Bigr)^{k+1}
\,,\qquad
 \gamma(\as) = \sum_{k=0}^{n-1} \gamma_k \Bigl(\frac{\as}{4\pi}\Bigr)^{k+1}
\,,\end{align}
while the QCD $\beta$ function is expanded as
\begin{align} \label{eq:beta_func}
 \beta(\as)
 \equiv \frac{\df\as}{\df\ln\mu}
 = -2 \as \sum_{k=0}^n \beta_k \Bigl(\frac{\as}{4\pi}\Bigr)^{k+1}
\,.\end{align}

\subsection{Roots of the \texorpdfstring{$\beta$}{beta} function}
\label{app:beta_roots}

We define the roots $\rt{i}$ of the $\beta$ function at N$^n$LL accuracy
according to \eq{expansion},
\begin{align}
 \beta(\as)
 = -8\pi a_s^2 \beta_n \prod_{i=1}^n \bigl(a_s - \rt{i} \bigr)
\,,\qquad
 a_s = \frac{\as}{4\pi}
\,.\end{align}
In addition to the double root at $\rt{0} \equiv 0$, one obtains the following roots:
\begin{alignat}{2} \label{eq:delta_i}
 \text{NLL}:& \qquad
 \rt{1} = -\frac{\beta_0}{\beta_1}
\,,\nn\\
 \text{NNLL}:& \qquad
 \rt{1,2} = \frac{-\beta_1 \pm \img \sqrt{4 \beta_0 \beta_2 - \beta_1^2 }}{2 \beta_2}
\,,\nn\\
 \text{N}^3\text{LL}:& \qquad
 \rt{i} = -\frac{1}{3 \beta_3} \left( \beta_2 + \xi^i C + \frac{\Delta_0}{\xi^i C}\right)
 \,,\qquad i = 1, 2, 3
\,.\end{alignat}
Note that at NNLL, $\rt{1,2}$ are imaginary valued for $n_f \le 5$.
The roots at N$^3$LL are expressed in terms of the constants
\begin{align}
 C = \sqrt[3]{\Delta_1 + \sqrt{\Delta_1^2 - \Delta_0^3}}
\,,\quad
 \Delta_0 = \beta_2^2 - 3 \beta_1 \beta_3
\,,\quad
 \Delta_1 = \beta_2^3 - \frac92 \beta_3 ( \beta_1 \beta_2 -  3 \beta_0 \beta_3 )
\,,\end{align}
and the three different roots are distinguished by different powers
of the primitive cube root of unity,
\begin{align}
 \xi = \frac{-1 + \sqrt{-3}}{2} = -\frac12 + \frac{\img}{2} \sqrt{3}
\,.\end{align}
This implies that one can take any cubic and square root in $C$.
For $n_f \le 7$, $\Delta_1^2 - \Delta_0^3 > 0$, such that $C$ is a real number
and one has two complex roots and one real root at N$^3$LL.

\subsection{Coefficients of the partial fraction decomposition}
\label{app:pfd_coefficients}

We require the partial fraction decompositions
\begin{align}
 \frac{\Gamma(\as)}{\beta(\as)}
 &= -\frac{1}{8\pi} \biggl[ \frac{\rs{\Gamma,0}}{a_s} + \sum_{i=1}^n \frac{\rs{\Gamma,i}}{a_s - \rt{i}} \biggr]
\,,\nn\\
 \frac{\gamma(\as)}{\beta(\as)}
 &= -\frac{1}{8\pi} \biggl[ \frac{\rs{\gamma,0}}{a_s} + \sum_{i=1}^n \frac{\rs{\gamma,i}}{a_s - \rt{i}} \biggr]
\,,\nn\\
 \frac{1}{\beta(\as)}
 &= -\frac{1}{8\pi} \left[ \frac{\barrsp{0}}{a_s^2} + \frac{\barrs{0}}{a_s} + \sum_{i=1}^n \frac{\barrs{i}}{a_s - \rt{i}}\right]
\,,\end{align}
where the roots $\rt{i}$ are given in \eq{delta_i}.
Note that the upper limit $n$ in the second sum
is governed by the perturbative order of the $\beta$ function,
even though $\gamma(\as)$ itself is truncated at one lower order.
Furthermore, only $1/\beta$ contains a double pole at $a_s = 0$.

The residues required at N$^n$LL are given by
\begin{alignat}{4} \label{eq:residues}
 \barrsp{0} &= \frac{1}{\beta_0}
 \,,\qquad
 &&\barrs{0} = -\frac{\beta_1}{\beta_0^2} \theta_{n>0}
 \,,\qquad
 &&\barrs{i} \,~= \frac{-2}{\beta'\bigl(4\pi\rt{i}\bigr)}
 \,,\quad
 &&i = 1, \dots, n
\,,\nn\\
 & &&\rs{\Gamma,0} = \frac{\Gamma_0}{\beta_0}
 \,,\qquad
 &&\rs{\Gamma,i}
 = \barrs{i} \, \Gamma\bigl(4\pi\rt{i}\bigr)
 \,,\quad
 && i = 1, \dots, n
\,,\nn\\
 & &&\rs{\gamma,0} = \frac{\gamma_0}{\beta_0} \theta_{n>0}
 \,,\qquad
 &&\rs{\gamma,i}
 = \barrs{i} \, \gamma\bigl(4\pi\rt{i}\bigr)
 \,,\quad
 && i = 1, \dots, n
\,.\end{alignat}
Note that the form of the residues at $a_s = 0$ is independent of the order $n$,
except that $\barrs{0}$ and $\rs{\gamma,0}$ only contribute beyond LL.
From \eq{anom_dims}, the cusp and noncusp anomalous dimensions at the appropriate
orders are given by
\begin{align}
 \Gamma\bigl(4\pi\rt{i}\bigr) = \sum_{k=0}^n \Gamma_k \rt{i}^{k+1}
\,,\qquad
 \gamma\bigl(4\pi\rt{i}\bigr) = \sum_{k=0}^{n-1} \gamma_k \rt{i}^{k+1}
\,,\end{align}
and the derivatives of the $\beta$ function can be evaluated using
\begin{align}
 \frac{1}{\barrs{i}}
 = \frac{\beta'\bigl(4\pi\rt{i}\bigr)}{-2} &
 = \sum_{k=0}^n (k+2) \beta_k \rt{i}^{k+1}
 = \rt{i}^2 \beta_n \prod_{\substack{j = 1\\ i \ne j}}^n \bigl(\rt{i} - \rt{j}\bigr)
\,.\end{align}

\addcontentsline{toc}{section}{References}
\bibliographystyle{jhep}
\bibliography{refs}

\end{document}